\documentclass[12pt]{article}
\usepackage{amssymb}
\usepackage{amsmath}
\usepackage{apacite}
\usepackage{latexsym}
\usepackage{epsfig}
\usepackage{graphicx}
\usepackage{subfigure}
\usepackage{wasysym}
\usepackage{threeparttable}
\usepackage{natbib}
\usepackage{color}
\usepackage{epstopdf}
\usepackage{bm}
\usepackage{float}
\usepackage{todonotes}
\usepackage{verbatim}

\usepackage{hyperref}

\def\boxit#1{\vbox{\hrule\hbox{\vrule\kern6pt
          \vbox{\kern6pt#1\kern6pt}\kern6pt\vrule}\hrule}}

\def\bse{\begin{eqnarray*}}
\def\ese{\end{eqnarray*}}
\def\be{\begin{eqnarray}}
\def\ee{\end{eqnarray}}
\def\bq{\begin{equation}}
\def\eq{\end{equation}}
\def\bse{\begin{eqnarray*}}
\def\ese{\end{eqnarray*}}

\begin{document}

\thispagestyle{empty} 
\baselineskip=28pt

\begin{center}
{\LARGE{\bf A General Bayesian Model for Heteroskedastic Data with Fully Conjugate Full-Conditional Distributions}}
%{\LARGE{\bf Just Google It: The Union of Deep Learning and Higher-order Spectral Analysis as an Answer to Nonlinear Time Series Classification}}

\end{center}

\baselineskip=12pt

%%
%%
%%
%%%%%%%%%%%%%%%%%%%%%%%%%%%%%%%%%%%%%%%%%%%%%%%%%%%%%%%%%%%%%%%%%%%%%%%%
\vskip 2mm
\begin{center}
Paul A. Parker\footnote{(\baselineskip=10pt to whom correspondence should be addressed)
Department of Statistics, University of Missouri,
146 Middlebush Hall, Columbia, MO 65211-6100, paulparker@mail.missouri.edu},
    Scott H. Holan\footnote{\baselineskip=10pt Department of Statistics, University of Missouri
146 Middlebush Hall, Columbia, MO 65211-6100, holans@missouri.edu}\,
%\footnote{\baselineskip=10pt U.S. Census Bureau, 4600 Silver Hill Road, Washington, D.C. 20233-9100, scott.holan@census.gov},
    and Skye A. Wills\footnote{\baselineskip=10pt USDA-Natural Resources Conservation Service, National Soil Survey Center, 100 Centennial Mall North, Room 152, Lincoln, NE 68508, USA}
\\
\end{center}
%
%
%
%
%%%%%%%%%%%%%%%%%%%%%%%%%%%%%%%%%%%%%%%%%%%%%%%%%%%%%%%%%%%%%%%%%%%%%%%%
\vskip 4mm
\baselineskip=12pt 
\begin{center}
{\bf Abstract}
\end{center}

Models for heteroskedastic data are relevant in a wide variety of applications ranging from financial time series to environmental statistics. However, the topic of modeling the variance function conditionally has not seen near as much attention as modeling the mean. Volatility models have been used in specific applications, but these models can be difficult to fit in a Bayesian setting due to posterior distributions that are challenging to sample from efficiently. In this work, we introduce a general model for heteroskedastic data. This approach models the conditional variance in a mixed model approach as a function of any desired covariates or random effects. We rely on new distribution theory in order to construct priors that yield fully conjugate full conditional distributions. Thus, our approach can easily be fit via Gibbs sampling. Furthermore, we extend the model to a deep learning approach that can provide highly accurate estimates for time dependent data. We also provide an extension for heavy-tailed data. We illustrate our methodology via three applications. The first application utilizes a high dimensional soil dataset with inherent spatial dependence. The second application involves modeling of asset volatility. The third application focuses on clinical trial data for creatinine.

\baselineskip=12pt
\par\vfill\noindent
\par\medskip\noindent
\clearpage\pagebreak\newpage \pagenumbering{arabic}
\baselineskip=24pt

\section{Introduction}\label{sec: intro}

Models for heteroskedastic data are relevant in a wide variety of applications ranging from finance \citep{engle1982autoregressive}, to ecology \citep{seekell2012conditional}, to political science \citep{alvarez1995american}, among others. These models tend to be specific in their application, and there does not seem to be a general approach for flexibly modeling the conditional variance of data. In contrast to this, there are quite general models for the conditional mean (generalized linear models and generalized linear mixed models). Furthermore, it is often desirable to incorporate various dependencies into models for heteroskedastic data.

A commonly used tool for modeling dependent data scenarios (spatial, temporal, etc.) is that of Bayesian hierarchical modeling. In the world of spatial and spatio-temporal statistics, Bayesian mixed models are often used with both covariates as well as spatial or spatio-temporal basis functions \citep{wikle2019spatio}. For Gaussian data, in general, these models focus on characterization of the mean function, often assuming a constant marginal variance. \citet{yan2007spatial} explore a spatially varying volatility model for lattice data, using a conditional autoregressive (CAR) structure for the variance. Their model fitting approach  requires the use of accept-reject steps within the MCMC sampling scheme in order to accommodate the non-conjugate full-conditional distributions induced by the Gaussian prior distribution, however, these approaches can be inefficient and difficult to tune, especially under high-dimensional settings. 

Volatility models have been studied more in depth in the time-series literature, beginning with the frequently used ARCH and GARCH models \citep{engle1982autoregressive, bollerslev1986generalized}. These approaches model volatility as a linear process. There have been numerous extensions of these models, but a common criticism of these approaches is that they require specific constraints on the regression parameters in order to constrain the volatility to be positive. A natural alterntaive is the EGARCH model \citep{nelson1991conditional}, which instead models the log volatility as a linear process, eliminating the need for parameter constraints. Bayesian inference for the EGARCH model has also been explored \citep{vrontos2000full}, however, this approach also uses non-conjugate Gaussian priors and Metropolis steps for sampling.

In a more general setting, \citet{johnson2018phenomenological} use a Gaussian process approach to model heteroskedastic data. However, their model is fit via maximum likelihood estimation and they do not consider Bayesian modeling or the accompanying computational concerns. \citet{baladandayuthapani2005spatially} use a general spline modeling approach to model spatially varying penalty terms for Gaussian prior distributions. Their approach is quite flexible, but relies on a high dimensional Metropolis-Hastings step to sample from the posterior distribution.

Our goal with this paper is to develop a general Bayesian modeling scheme for heteroskedastic data that can flexibly model both the mean and variance functions. Importantly, this approach relies on new distribution theory in order to yield fully conjugate full-conditional distributions, which can be efficiently sampled from via Gibbs sampling. The flexibility of this model allows for application in a variety of domains, including dependent data scenarios, which we illustrate through both a spatial and a temporal example.

The outline of the remainder of this paper follows. In Section~\ref{sec: methods} we present our general modeling approach, as well as the relevant distribution theory. In addition to the general model, we also include and extension to deep learning and an extension to heavy-tailed data. We illustrate with an application to soil scoring curves in Section~\ref{sec: soil}, an application to asset volatility modeling in Section~\ref{sec: finance}, as well as an application to creatinine analysis with heavy tails in Section~\ref{sec: creatinine}. Finally, we provide concluding remarks as well as discussion in Section~\ref{sec: disc}.
	
\section{Methodology}\label{sec: methods}

Our general approach is to use Bayesian hierarchical modeling to model both the conditional mean and the conditional variance for Gaussian data. This type of modeling for the mean is quite common, however some innovation is required for adequate modeling of the variance. In particular, we rely on new distribution theory as well as an appropriate link function to construct priors that yield conjugate full-conditional distributions. In this section, we present the relevant distribution theory and then present our new modeling approach.

\subsection{Multivariate Log-Gamma Distribution}

The cornerstone of our modeling framework is the Multivariate Log-Gamma distribution (MLG), introduced by \citet{bradley2018computationally} and \citet{bradley2019bayesian} in order to model dependent data using a Poisson likelihood. The density for the MLG distribution is given as 
\begin{equation}
    \hbox{det}(\bm{V}^{-1})
    \left\{ \prod_{i=1}^n \frac{\kappa_i^{\alpha_i}}{\Gamma(\alpha_i)}\right\}
    \hbox{exp}\left[\bm{\alpha}' \bm{V}^{-1}(\bm{Y- \mu}) -
    \bm{\kappa}' \hbox{exp}\left\{\bm{V}^{-1}(\bm{Y- \mu}) \right\}\right],
\end{equation} denoted by $\hbox{MLG}(\bm{\mu}, \mathbf{V}, \bm{\alpha}, \bm{\kappa})$. Sampling from $\bm{Y}\sim \hbox{MLG}(\bm{\mu}, \mathbf{V}, \bm{\alpha}, \bm{\kappa})$ is straightforward using the following steps:
\begin{enumerate}
        \item Generate a vector $\mathbf{g}$ as $n$ independent Gamma random variables with shape $\alpha_i$ and rate $\kappa_i$, for $i=1,\ldots,n$
        \item Let $\mathbf{g}^*=\hbox{log}(\mathbf{g})$
        \item Let $\mathbf{Y}=\mathbf{V g}^* + \bm{\mu}$.
    \end{enumerate} 

Bayesian inference using the MLG prior distribution also often requires simulation from the conditional multivariate log-Gamma distribution ($\hbox{cMLG}$). Letting $\mathbf{Y} \sim \hbox{MLG}(\bm{\mu},  \mathbf{V}, \bm{\alpha}, \bm{\kappa})$, \citet{bradley2018computationally} show that $\mathbf{Y}$ can be partitioned into $(\mathbf{Y_1}', \mathbf{Y_2}')'$, where $\mathbf{Y_1}$ is $r$-dimensional and $\mathbf{Y_2}$ is $(n-r)$-dimensional. The matrix $\mathbf{V}^{-1}$ is also partitioned into $\left[\mathbf{H \; B}  \right]$, where $\mathbf{H}$ is an $n \times r$ matrix and $\mathbf{B}$ is an $ n \times (n - r)$ matrix. Then 
$$\bm{Y_1} | \bm{Y_2} = \bm{d}, \bm{\mu}^*, \bm{H}, \bm{\alpha}, \bm{\kappa} \sim \hbox{cMLG}(\bm{\mu}^*, \bm{H}, \bm{\alpha}, \bm{\kappa})$$ with density
\begin{equation}
    M \hbox{exp} \left\{\bm{\alpha}' \bm{H Y_1} 
     - \bm{\kappa}' \hbox{exp}(\bm{H Y_1} - \bm{\mu}^*)\right\} I\left\{(\bm{Y_1}' , \bm{d}')' \in \mathcal{M}^n \right\},
\end{equation} where $\bm{\mu}^*=\mathbf{V}^{-1}\bm{\mu} - \mathbf{Bd}$, and $\mathit{M}$ is a normalizing constant.  \citet{bradley2018computationally} show that it is also straightforward to sample from the cMLG distribution using $(\mathbf{H}'\mathbf{H})^{-1}\mathbf{H}'\mathbf{Y}$, where $\mathbf{Y}$ is sampled from $\hbox{MLG}(\bm{\mu}, \mathbf{I}, \bm{\alpha}, \bm{\kappa})$.

Another important result given by \citet{bradley2018computationally} is that $\hbox{MLG}(\mathbf{c}, \alpha^{1/2}\mathbf{V}, \alpha \mathbf{1}, \alpha \mathbf{1})$ converges in distribution to a multivariate normal distribution with mean $\mathbf{c}$ and covariance matrix $\mathbf{V}$ as the value of $\alpha$ approaches infinity. This allows for the use of MLG priors in place of Gaussian priors, in situations where it is computationally preferable, while still achieving the same posterior in the limit of $\alpha$.

\subsection{General Bayesian Heteroskedasticity Model}

We now present our General Bayesian Heteroskedasticity Model (GBHM), which makes use of MLG prior distributions to model the variance function,

\begin{equation}
    \begin{aligned}
        y_i | \mu_i, \sigma_i^2 & \stackrel{ind.}{\sim} \hbox{N}(\mu_i, \sigma^2_i), \; i=1,\ldots,n \\
        \mu_i &= \bm{x}_{1i}'\bm{\beta}_1 + \bm{\psi}_{1i}'\bm{\eta}_1 \\
        -\hbox{log}(\sigma_i^2) &= \bm{x}_{2i}'\bm{\beta}_2 + \bm{\psi}_{2i}'\bm{\eta}_2 \\
        \bm{\eta}_1 & \sim \hbox{N}(\bm{0}, \sigma^2_{\eta_1}\bm{\hbox{I}}) \\
        \bm{\eta}_2 & \sim \hbox{MLG}(\bm{0}, \alpha^{1/2} \sigma^2_{\eta_2} \bm{\hbox{I}}, \alpha \bm{1}, \alpha\bm{1}) \\
        \bm{\beta}_1 & \sim \hbox{N}(\bm{0}, \sigma^2_{\beta_1}\bm{\hbox{I}}) \\
        \bm{\beta}_2 & \sim \hbox{MLG}(\bm{0}, \alpha^{1/2} \sigma^2_{\beta_2} \bm{\hbox{I}}, \alpha \bm{1}, \alpha\bm{1}) \\
        \sigma^2_{\eta_1} & \sim \hbox{IG}(a, b) \\
        \frac{1}{\sigma_{\eta_2}} & \sim \hbox{Log-Gamma}^+(\omega, \rho) \\
        & \sigma^2_{\beta_1}, \sigma^2_{\beta_2}, \alpha, a, b, \omega, \rho > 0.
    \end{aligned}
\end{equation} 
We use a Normal data likelihood for observations $i=1,\ldots,n$, with unknown mean, $\mu_i$, and variance, $\sigma^2_i$. The mean function is modeled as a linear combination using a traditional mixed effects framework, where $\bm{x}_{1i}$ is a $p_1$-dimensional set of fixed effects covariates and $\bm{\psi}_{1i}$ is an $r_1$-dimensional set of random effects covariates. In many cases, it may be desirable to incorporate basis expansions into $\bm{\psi}_{1i}$ (e.g. spatial settings). We use a conjugate Normal prior distribution for the elements of $\bm{\beta}_1$, with mean zero and a fixed variance of $\sigma^2_{\beta_1}$. For the elements of $\bm{\eta}_1$, we also use a Normal prior, with mean zero, though with a common but unknown variance, $\sigma^2_{\eta_1}$. We then place an Inverse Gamma prior distribution on $\sigma^2_{\eta_1}$, as commonly done in mixed effects frameworks. 

We also model the variance function of the data through a linear combination of covariates. Importantly, we use the negative log link function here, which yields conjugate full conditional distributions when paired with the MLG prior distribution. Note that this is the same link function used by \citet{baladandayuthapani2005spatially}, however, their use of Normal priors resulted in a much more difficult sampling approach. The $p_2$-dimensional vector 
$\bm{x}_{2i}$ and the $r_2$-dimensional vector $\bm{\psi}_{2i}$ may be specified to be equivalent to $\bm{x}_{1i}$ and $\bm{\psi}_{1i}$, though we allow them to differ here for flexibility in modeling. The MLG priors placed on $\bm{\beta}_2$ and $\bm{\eta}_2$ are asymptotically equivalent to a Normal prior, however the MLG specification results in conjugate full-conditional distributions which may be easily sampled from. Alternatively, one could further model the parameters of the MLG prior for extra flexibility. Similar to $\bm{\eta}_1$, the elements of $\bm{\eta}_2$ share a common prior variance that is unknown, $\sigma^2_{\eta_2}$. Finally, we place a Log-Gamma prior truncated below at zero on $\frac{1}{\sigma_{\eta_2}}$. This prior was chosen to yield a conjugate full conditional distribution for $\sigma^2_{\eta_2}$.  The posterior distribution can be sampled from using a Gibbs sampling approach with the following full conditional distributions:
\begin{align*}
    \bm{\beta}_1 | \cdot & \sim \hbox{N}_{p_1}\left( 
        \bm{\mu}=(\bm{X}_1'\bm{\Sigma}_y\bm{X}_1 + \frac{1}{\sigma^2_{\beta_1}} \bm{I}_{p_1})^{-1} \bm{X}_1'\bm{\Sigma}_y(\bm{y} - \bm{\Psi}_1\bm{\eta}_1),
        \bm{\Sigma} = (\bm{X}_1'\bm{\Sigma}_y\bm{X}_1 + \frac{1}{\sigma^2_{\beta_1}} \bm{I}_{p_1})^{-1}
    \right) \\
        \bm{\eta}_1 | \cdot & \sim \hbox{N}_{r_1}\left( 
        \bm{\mu}=(\bm{\Psi}_1'\bm{\Sigma}_y\bm{\Psi}_1 + \frac{1}{\sigma^2_{\eta_1}} \bm{I}_{r_1})^{-1} \bm{\Psi}_1'\bm{\Sigma}_y(\bm{y} - \bm{X}_1\bm{\beta}_1),
        \bm{\Sigma} = (\bm{\Psi}_1'\bm{\Sigma}_y\bm{\Psi}_1 + \frac{1}{\sigma^2_{\eta_1}} \bm{I}_{r_1})^{-1}
    \right) \\
    \bm{\beta}_2 | \cdot &\sim \hbox{cMLG}(\bm{H}_{\beta_2}, \bm{\alpha}_{\beta_2}, \bm{\kappa}_{\beta_2}) \\
    & \bm{H}_{\beta_2} = \left[
                            \begin{array}{c}
                            \bm{X}_2  \\
                            \alpha^{-1/2} \frac{1}{\sigma_{\beta_2}} \bm{I}_{p_2} 
                            \end{array}
                            \right], \quad 
        \bm{\alpha}_{\beta_2} = \left(\frac{1}{2} \bm{1}_n', \alpha \bm{1}_{p_2}'\right)', \quad \\
        & \bm{\kappa}_{\beta_2} = \left(\frac{1}{2}\left\{(\bm{y}-\bm{\mu}_y)^2 \odot \hbox{exp}(\bm{\Psi}_2 \bm{\eta}_2)\right\}', \alpha \bm{1}_{p_2}'\right)' \\
                \bm{\eta}_2 | \cdot &\sim \hbox{cMLG}(\bm{H}_{\eta_2}, \bm{\alpha}_{\eta_2}, \bm{\kappa}_{\eta_2}) \\
        & \bm{H}_{\eta_2} = \left[
                            \begin{array}{c}
                            \bm{\Psi}_2  \\
                            \alpha^{-1/2} \frac{1}{\sigma_{\eta_2}} \bm{I}_{r_2} 
                            \end{array}
                            \right], \quad 
        \bm{\alpha}_{\eta_2} = \left(\frac{1}{2} \bm{1}_n', \alpha \bm{1}_{r_2}'\right)', \quad \\
        & \bm{\kappa}_{\eta_2} = \left(\frac{1}{2}\left\{(\bm{y}-\bm{\mu}_y)^2 \odot \hbox{exp}(\bm{X}_2 \bm{\beta}_2)\right\}', \alpha \bm{1}_{r_2}'\right)' \\
        \sigma^2_{\eta_1} | \cdot & \sim \hbox{IG}\left(a + \frac{r_1}{2}, \; b + \frac{\bm{\eta}_1'\bm{\eta}_1}{2} \right) \\
        \frac{1}{\sigma_{\eta_2}} | \cdot & \sim \hbox{cMLG}(\bm{H}_{\sigma}, \bm{\omega}_{\sigma}, \bm{\rho}_{\sigma}) \times I(\sigma_{\eta_2} > 0) \\
        & \bm{H}_{\sigma} = (\alpha^{-1/2} \bm{\eta}_2', 1)' \quad
        \bm{\omega}_{\sigma} = (\alpha \bm{1}_{r_2}', \omega)' \quad 
        \bm{\rho}_{\sigma} = (\alpha \bm{1}_{r_2}', \rho)'.
\end{align*}The derivations of these full conditional distributions may be found in the Appendix.

This modeling framework allows for a high degree of flexibility in modeling both the mean and variance of the data. The GBHM contains two important particular cases. When $\bm{X}_2$ contains an intercept only, and $\bm{\Psi}_2$ is empty, the results is a constant variance model (i.e. a basic Gaussian mixed model). In fact, under this intercept only model, use of a univariate log Gamma prior distribution for $\beta_0$ is equivalent to the commonly used conjugate inverse Gamma prior for the variance. When $\bm{X}_2$ contains discrete covariates only,  and $\bm{\Psi}_2$ is empty, the result is a group level variance model.

\subsection{BGHM for Heavy-Tailed Data}

 In many applications, such as financial time series, heavy tailed data models are used in place of the traditional Normal distribution \citep{hall2003inference}. One such distribution is the Laplace, or Double Exponential distribution, having density
$$
\frac{1}{2b} \hbox{exp}\left (\frac{|y-\mu|}{b} \right), \; b>0,
$$ with mean $\mu$ and variance equal to $2b^2$.

We are able to leverage the fact that the Laplace distribution can be written as a scale mixture of Exponential distributions in order to extend our BGHM to a heavy tailed data setting while still retaining conjugacy. In particular, we make use of the identity 
$$
\frac{1}{2b} \hbox{exp}\left (\frac{|y|}{b} \right) = \int_0^{\infty} \frac{1}{\sqrt{2 \pi s}} \hbox{exp}\left({-\frac{y^2}{2s}}\right) \frac{1}{2b^2} \hbox{exp} \left({-\frac{s}{2b^2}} \right) ds
$$ in order to construct a data augmentation approach to sampling from the posterior distribution under the BGHM with a Laplace rather than Gaussian data model. This identity implies that we may model the Laplace data as Gaussian conditional on the variance, and then place an Exponential prior on the Gaussian variance with mean $2b^2$, which is the variance of the Laplace distribution. By further modeling the mean of this Exponential prior, we can directly model the variance of the Laplace data distribution. Similar data augmentation approaches have been explored for both prior distributions on regression parameters \citep{park2008bayesian} as well as for homoskedastic Laplace data \citep{choi2013analysis}.

Recognizing this scale mixture identity, we write the Laplace BGHM as

\begin{equation}
    \begin{aligned}
        y_i | \mu_i, s_i & \stackrel{ind.}{\sim} \hbox{N}(\mu_i, s_i), \; i=1,\ldots,n \\
                s_i | \sigma^2_i & \stackrel{ind}{\sim} \hbox{Exp}(\sigma_i^2) \\
        \mu_i &= \bm{x}_{1i}'\bm{\beta}_1 + \bm{\psi}_{1i}'\bm{\eta}_1 \\
        -\hbox{log}(\sigma_i^2) &= \bm{x}_{2i}'\bm{\beta}_2 + \bm{\psi}_{2i}'\bm{\eta}_2 \\
        \bm{\eta}_1 & \sim \hbox{N}(\bm{0}, \sigma^2_{\eta_1}\bm{\hbox{I}}) \\
        \bm{\eta}_2 & \sim \hbox{MLG}(\bm{0}, \alpha^{1/2} \sigma^2_{\eta_2} \bm{\hbox{I}}, \alpha \bm{1}, \alpha\bm{1}) \\
        \bm{\beta}_1 & \sim \hbox{N}(\bm{0}, \sigma^2_{\beta_1}\bm{\hbox{I}}) \\
        \bm{\beta}_2 & \sim \hbox{MLG}(\bm{0}, \alpha^{1/2} \sigma^2_{\beta_2} \bm{\hbox{I}}, \alpha \bm{1}, \alpha\bm{1}) \\
        \sigma^2_{\eta_1} & \sim \hbox{IG}(a, b) \\
        \frac{1}{\sigma_{\eta_2}} & \sim \hbox{Log-Gamma}^+(\omega, \rho) \\
        & \sigma^2_{\beta_1}, \sigma^2_{\beta_2}, \alpha, a, b, \omega, \rho > 0,
    \end{aligned}
\end{equation} 
 where $\hbox{Exp}(a)$ represents the Exponential distribution with mean $a$. We note that this is a slightly different parameterization than the one explored by \citet{choi2013analysis}, and was chosen in order to model the variance (i.e. $\sigma^2_t$) directly as done in the  Gaussian ESVM. 

Sampling from the posterior distribution under the Laplace BGHM can be done using Gibbs sampling. The only full conditional distributions that differ from the Gaussian ESVM are $s_t$, $\bm{\beta}_2$, and $\bm{\eta}$. The data augmentation variables can be sampled using the full conditional distribution $\frac{1}{s_i} | \cdot  \stackrel{ind}{\sim} \hbox{Inverse Gaussian}\left(\mu_s = \left\{\frac{1}{(y_i-\mu_i)^2\sigma^2_i} \right\}^{1/2}, \; \lambda_s=\frac{2}{\sigma^2_i} \right)$ for $i=1,\ldots,n$. We also have $\bm{\beta_2} | \cdot \sim \hbox{cMLG}(\bm{H}_{\beta_2}, \bm{\alpha}_{\beta_2}, \bm{\kappa}_{\beta_2})$ and $\bm{\eta}_2 | \cdot \sim \hbox{cMLG}(\bm{H}_{\eta_2}, \bm{\alpha}_{\eta_2}, \bm{\kappa}_{\eta_2}),$ where $\bm{H}_{\beta_2} = \left[
                            \begin{array}{c}
                            \bm{X}  \\
                            \alpha^{-1/2} \frac{1}{\sigma_{\beta_2}} \bm{I}_{p_2} 
                            \end{array}
                            \right],$
                            $\bm{H}_{\eta_2} = \left[
                            \begin{array}{c}
                            \bm{\Psi}  \\
                            \alpha^{-1/2} \frac{1}{\sigma_{\eta_2}} \bm{I}_{r_2} 
                            \end{array}
                            \right],$ $\bm{\alpha}_{\beta_2} = \left( \bm{1}_n', \alpha \bm{1}_{p_2}'\right)'$, $\bm{\alpha}_{\eta_2} = \left( \bm{1}_n', \alpha \bm{1}_{r_2}'\right)',$ 
          $\bm{\kappa}_{\beta_2} = \left(\left\{\bm{s}\odot\hbox{exp}(\bm{\Psi}_2\bm{\eta}_2)\right\}, \alpha \bm{1}_{p_2}'\right)',$ and $\bm{\kappa}_{\eta_2} = \left(\left\{\bm{s}\odot\hbox{exp}(\bm{X}_2\bm{\beta}_2)\right\}, \alpha \bm{1}_{r_2}'\right)'$. We provide these derivations in the Appendix.

\subsection{Echo State Volatility Model}

The generality of the GBHM allows it to be extended in many ways. One extension that we focus on here is the use of deep modeling techniques. In particular, recurrent neural networks (RNNs) have been used extensively for prediction of time series data. Traditionally, these neural networks are fit with stochastic gradient descent techniques in order to minimize some loss function, typically mean squared error.  One alternative to traditional RNNs is the Echo State Network (ESN), which randomly samples and fixes the hidden layer weights before model fitting \citep{jaeger2007echo}. The ESN for a univariate time series can be written generally as
\begin{equation}
    \begin{aligned}
        y_t &= g_o(\bm{h}_t'\bm{\eta}) \\
        \bm{h}_t &= g_h(\bm{W}\bm{h}_{t-1} + \bm{U}\bm{x}_t),
    \end{aligned}
\end{equation} where $y_t$ is the value of the time series at time $t=1,\ldots,T$. The $n_h$-dimensional hidden layer at time $t$ is represented by $\bm{h}_t$, and is constructed using $\bm{h}_{t-1}$ as well as a $p$-dimensional vector time indexed covariates, $\bm{x}_t$. The $n_h \times n_h$ dimensional weight matrix $\bm{W}$ induces the recurrent component of the model and the $n_h \times p$ dimensional weight matrix acts on the covariates.  The activation function, $g_h(\cdot)$, allows for nonlinearity in the modeling approach. We will use the hyperbolic tangent function, although other activation functions could be explored. Finally, the output layer activation function is given by $g_o(\cdot)$. For a continuous valued time series, this will be the identity function. Importantly, the weight matrices $\bm{W}$ and $\bm{U}$ are randomly chosen and fixed. Thus, the only parameters that must be learned are $\bm{\eta}$. This results in a much easier optimization problem, for which typical linear regression techniques may be used in the case of continuous valued time series. Alternatively, it may be desirable to add regularization in order to prevent overfitting, through the use of penalized regression approaches.

More recently, \citet{mcdermott2017ensemble} used the ESN in a statistical context by linking the hidden layer outputs to a likelihood. They quantify uncertainty through the use of an ensemble of ESN models. \citet{mcdermott2019bayesian} provide a natural alternative for uncertainty quantification through a Bayesian ESN by placing a prior distribution on the output layer weights. Finally, \citet{mcdermott2019deep} consider deeper model hierarchies through the use of multiple hidden layers. Although these approaches are the first use of the ESN in a statistical context for uncertainty quantification, they only consider the ESN as a means to model the mean function.

We combine the concept of the ESN with our GBHM hierarchy in order to model temporal volatility through a nonlinear deep learning lens. This model is termed the Echo State Volatility Model (ESVM) and written as
\begin{equation}
    \begin{aligned}
        y_t | \mu, \sigma_t^2 & \stackrel{ind.}{\sim} \hbox{N}(\mu, \sigma^2_t), \; t=1,\ldots,T \\
        -\hbox{log}(\sigma_t^2) &=  \bm{h}_t'\bm{\eta} \\
        \bm{h}_t &= g_h(\bm{W}\bm{h}_{t-1} + \bm{U}\bm{x}_t) \\
        \bm{\eta} & \sim \hbox{MLG}(\bm{0}, \alpha^{1/2} \sigma^2_{\eta} \bm{\hbox{I}}, \alpha \bm{1}, \alpha\bm{1}) \\
        \frac{1}{\sigma_{\eta}} & \sim \hbox{Log-Gamma}^{(c,\infty)}(\omega, \rho) \\
        \mu & \sim \hbox{N}(0, \sigma^2_{\beta}) \\
        & \sigma^2_{\beta}, \alpha, \omega, \rho, c > 0.
    \end{aligned}
\end{equation} 
The elements of $\bm{W}$ and $\bm{U}$ are randomly generated independently from a Normal distribution with mean zero and standard deviation 0.1. We found that our results were not very sensitive to this choice, but depending on the application it may be desirable to use cross validation to select this distribution. Additionally, after generating $\bm{W}$, we scale by $\frac{\delta}{|\lambda_w|}$, as recommended by \citet{mcdermott2019bayesian} to limit the spectral radius of $\bm{W}$ and prevent unstable behavior. Here, we set $\delta=0.1$ and $\lambda_w$ is the largest eigenvalue of $\bm{W}$. The prior distributions used here resemble those of the BGHM, with the exception of $\frac{1}{\sigma_{\eta}}$. Here we truncate below at  $c$ rather than zero, where we set $c=7$ for this work. Regularization is a critical component of deep learning approaches, so we make this decision to prevent large values of the prior variance. Alternatively, one could place a discrete uniform prior on the prior variance, or use another prior entirely, though conjugate sampling may not be retained in the latter case.

Importantly, because the matrix $\bm{A}$ is randomly generated and considered fixed before model fitting, the ESVM can be seen as a special case of the BGHM. Thus, the same Gibbs sampling approach used for the BGHM can be used in this case as well. In addition, a Laplace version of the ESVM may be used just as in the BGHM case.

\section{Application to Soil Characterization}\label{sec: soil}

Characterization of various soil health indicators is an important topic in the soil science literature. For example, conditional cumulative density functions may be used to assign health scores for various metrics \citep{fine2017statistics}. The distribution of soil measurements can vary according to different soil types, thus soil scoring approaches rely on heteroskedastic data models, however current approaches only consider variance as a function of discrete covariates. In addition to soil health, soil carbon stock measurements are used for carbon accounting, verification, and assessment \citep{sanderman2010accounting}. Thus, sufficient characterization of soil carbon stock can have important consequences in regards to mitigation of climate change \citep{smith2020measure}. The development of the GBHM allows for variability of soil characteristics such as carbon to be modeled as a function of discrete as well as continuous covariates. Furthermore, through the use of spatial basis functions, the variance function for these soil characteristics may even be modeled spatially.

\subsection{USDA-NRCS Rapid Carbon Data}

\begin{figure}[H]
    \begin{center}
        \includegraphics[width=150mm]{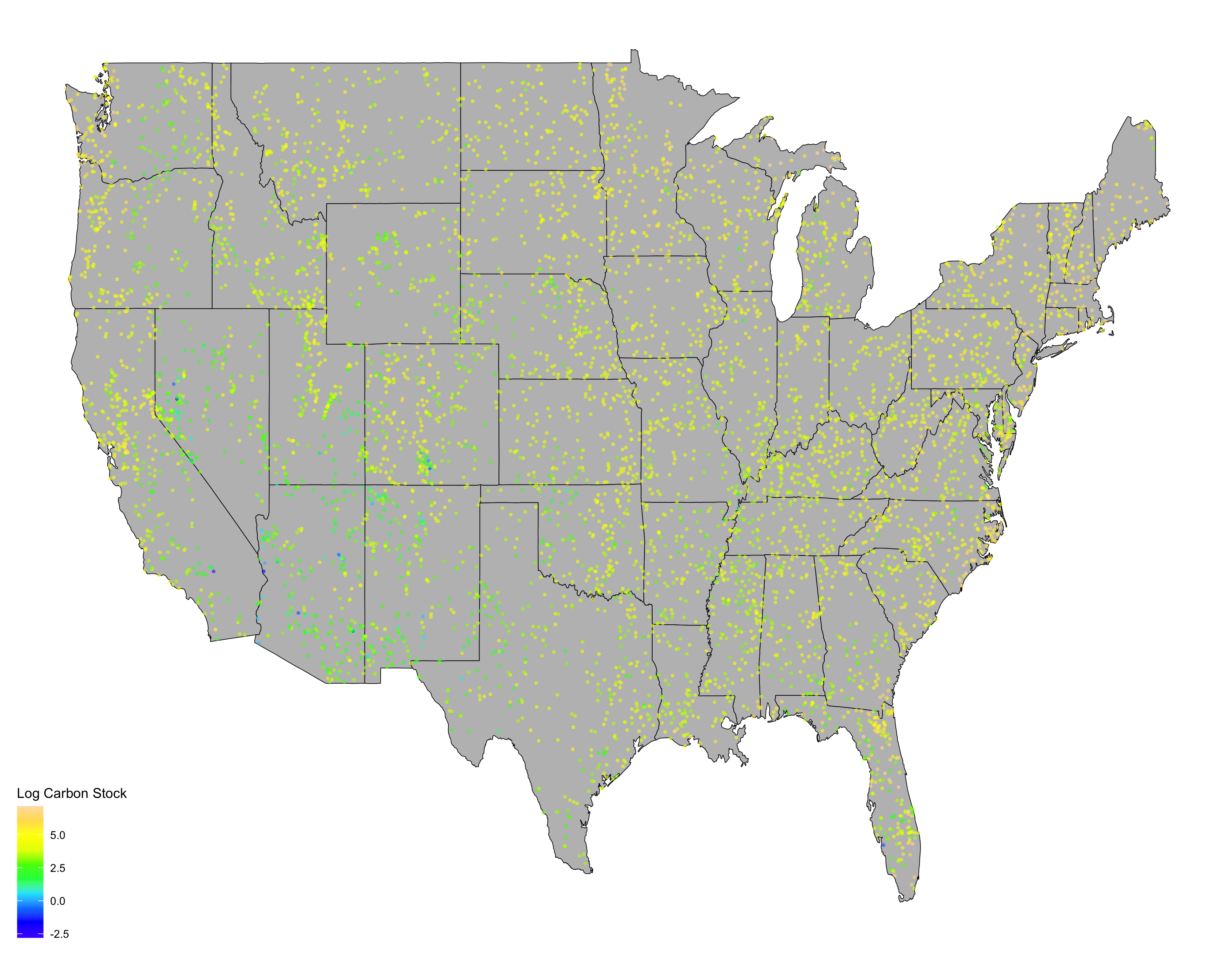}
         \caption{\baselineskip=10pt Map of soil carbon stock at 30 cm on the log scale.}
         \label{fig: datamap}
    \end{center}               
\end{figure}

As an illustration of our GBHM, we consider the USDA-Natural Resources Conservation Service (USDA-NRCS)  Rapid Carbon Data \citep{NRCSsoil}. This dataset consists of over 6,000 observations, with measurements including soil carbon stock at 30 cm, mean annual temperature, annual precipitation, soil order category, as well as longitude and latitude. Figure \ref{fig: datamap} shows the locations as well as the log carbon stock. The data was primarily collected in 2010. Importantly, this dataset spans spatial locations across the entire continental United States. 

\subsection{Carbon Stock Analysis}

We consider log carbon stock as a response variable,$y_i$, and seek to characterize the distribution under various subsets of covariates. In particular, we fit the GBHM  with three different settings.
\begin{enumerate}
    \item The vector $\bm{x}_{1i}$ contains soil order (discrete), as well as temperature and precipitation (continuous). The vector $\bm{x}_{2i}$ contains only an intercept (i.e. a constant variance model). The vectors $\bm{\psi}_{1i}$ and $\bm{\psi}_{2i}$ are empty.
    \item The vectors $\bm{x}_{1i}$ and $\bm{x}_{2i}$ both contain soil order (discrete), as well as temperature and precipitation (continuous). The vectors $\bm{\psi}_{1i}$ and $\bm{\psi}_{2i}$ are empty.
    \item The vectors $\bm{x}_{1i}$ and $\bm{x}_{2i}$ both contain soil order (discrete), as well as temperature and precipitation (continuous). The vectors $\bm{\psi}_{1i}$ and $\bm{\psi}_{2i}$ both contain a set of spatial basis functions.
\end{enumerate}
The basis functions in model 3 were bisquare basis functions automatically selected through the use of the \texttt{FRK} package in R \citep{frk}. These were selected at two resolutions, resulting in 148 total basis functions. In every case, we use the weakly informative prior, $\sigma^2_{\beta_1}=\sigma^2_{\beta_1}=\alpha=\omega=\rho=1000$ and $a=b=0.5$. MCMC was run for 5,000 iterations, where the first 1,000 iterations were discarded as burn-in. We assessed convergence through traceplots of the chains, and found no evidence of lack of convergence.

We evaluate the three models using three separate criteria. First, we calculate the deviance information criterion (DIC) and Watanabe-Akaike information criterion (WAIC) \citep{watanabe2010asymptotic} for each model fit using the entire dataset. This gives an in sample view of goodness of fit, where a lower value of DIC or WAIC indicates a better fit. Although WAIC is calculated in sample, it is used to approximate the out of sample expected predictive density. Second, we use five-fold cross validation to calculate the out of sample mean squared error between the residuals and the estimated variance,
$$
MSEV = \frac{1}{n}\sum_{i=1}^n \left((y_i - \hat{\mu}_i)^2 - \hat{\sigma}^2_i  \right)^2,
$$
where $\hat{\mu}_i$ and $\hat{\sigma}^2_i$ are the posterior mean point estimates. This metric is chosen based on the definition of the variance as the expected value of $(y_i-\mu_i)^2$. A summary of these results can be found in Table \ref{tab: soil}. Model~1 performs the worst under both DIC and MSEV, giving clear indication that a constant variance model is not sufficient to characterize soil carbon stock. Model~2 is able to improve upon this by considering the covariates in the model for the variance function. Finally, Model~3 outperforms both other models by including additional spatial dependence structure in both the mean and variance functions. In particular, Model 3 is able to provide a substantial 29\% reduction in MSEV when compared to the constant variance model. 

\begin{table}[H]
\begin{center}
 \begin{tabular}{||c | c c c ||} 
 \hline
 Estimator & DIC & WAIC & MSEV  \\ [0.5ex] 
 \hline\hline
 Model 1 & $1.57 \times 10^{4}$ &  $1.57 \times 10^{4}$ & $2.24$  \\ 
 \hline
 Model 2 & $1.53 \times 10^{4}$ &  $1.59 \times 10^{4}$ & $2.17$  \\
 \hline
 Model 3 & $\bm{1.39 \times 10^{4}}$ & $\bm{1.52 \times 10^{4}}$ & $\bm{1.59}$   \\ [1ex] 
 \hline
\end{tabular}
\caption{DIC, WAIC, and MSEV for each of the three models fit using the USDA-NRCS Rapid Carbon soil data. DIC and WAIC are calculated by fitting the model on the full dataset, while MSEV is calculated using five-fold cross validation.}
\label{tab: soil}
\end{center}
\end{table}

\section{Application to Asset Volatility}\label{sec: finance}
Models for heteroskedastic data are critical in the realm of financial time series. In particular, asset volatility is an important component of options pricing \citep{black1973pricing}, thus sufficient modeling of asset volatility constitutes an important problem. Both ARCH and GARCH \citep{engle1982autoregressive, bollerslev1986generalized} are two common approaches to this problem that model the conditional variance as a linear process. These approaches require certain boundary conditions on the model parameter to ensure a positive variance, which can make posterior sampling computationally difficult in a Bayesian setting. As an alternative, the GBHM may be used to fit a Bayesian EGARCH model \citet{nelson1991conditional} with Gibbs sampling. However still, nonlinear modeling techniques may be necessary to capture the dynamics of asset volatility. Thus, we propose the use of the ESVM to model asset volatility, which allows for very flexible nonlinear modeling of the variance function, with temporal dependence, while still being computationally efficient.

As an illustration of the ESVM, we consider Dow Jones Industrial Average daily close prices from December 2015 through December 2018. We use the log daily returns as a response variable, in order to model volatility. In addition, we obtain the Chicago Boards Options Exchange Volatility Index (VIX) over this same time period, for potential use as a covariate.

\begin{figure}[H]
    \begin{center}
        \includegraphics[width=150mm]{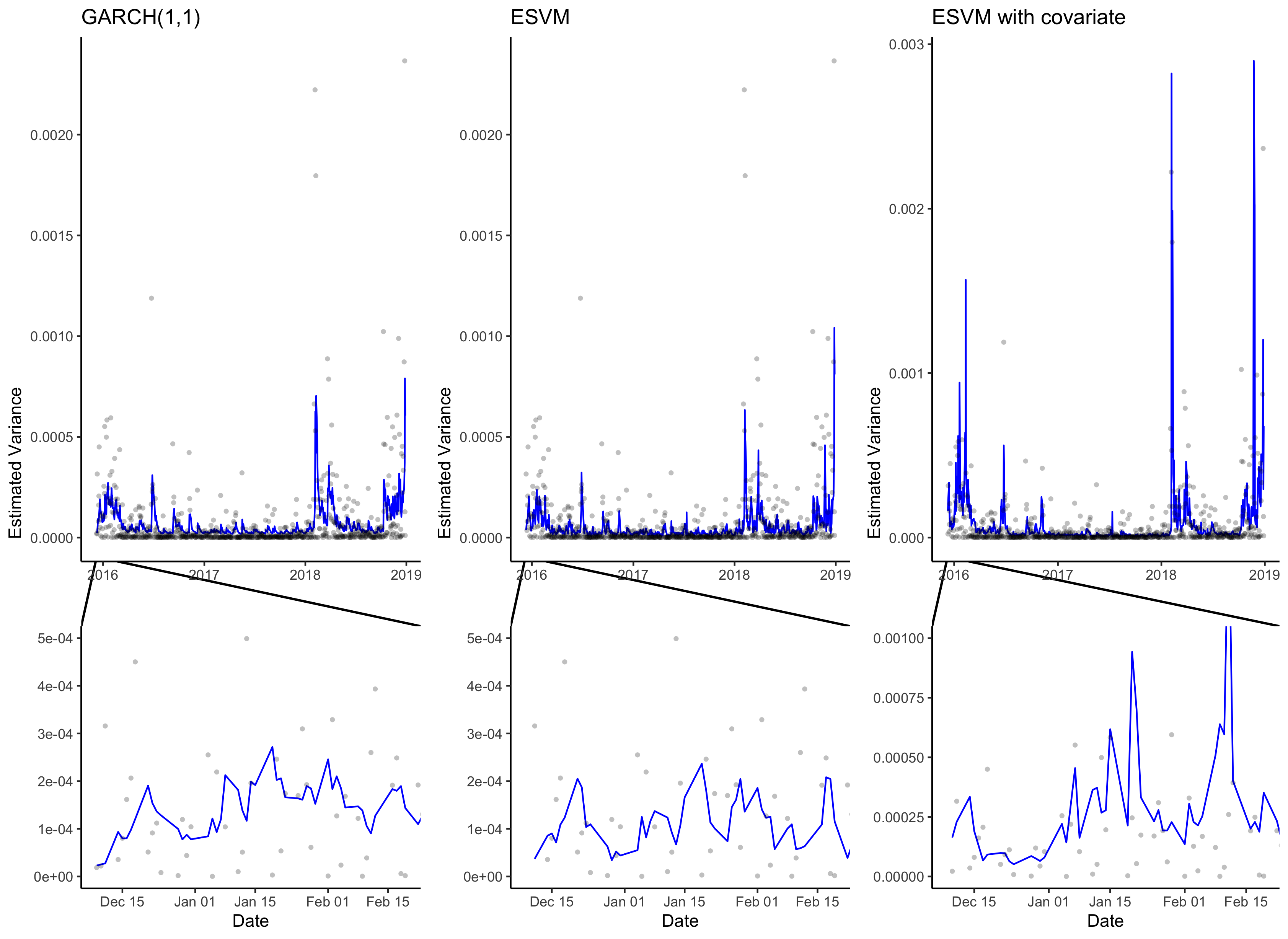}
         \caption{\baselineskip=10pt Plot of estimated volatility under each model when fit to the Dow Jones data. Points represent the squared log returns. The lower plots zoom in on the time period from December 2015 to February 2016.}
         \label{fig: volatility}
    \end{center}               
\end{figure}

We fit the ESVM on the Dow Jones data using $\alpha=\omega=\rho=1000.$ We let $\bm{x}_t=\left(1, \hbox{log}(y_{t-1}^2)\right)$. We also fit a separate ESVM using log VIX as a covariate for which $\bm{x}_t=\left(1, \hbox{log}(y_{t-1}^2), \hbox{log}(\hbox{VIX}_t) \right)$. Finally, we compare the two ESVM approaches to a GARCH(1,1) model, which is commonly used in volatility applications. The GARCH model is fit using Stan \citep{carpenter2017stan}. We explored the Laplace ESVM as well, but found that it did not provide any additional benefit for this particular dataset. Figure~\ref{fig: volatility} shows the fits for each of these models as well as the squared log returns for reference. In addition, we zoom in on a shorter time frame to give insight into the short term behavior of each model. In general, all three approaches seem to capture the same temporal pattern. In particular, they indicate high periods of volatility at the beginning and end of 2018 as well as the beginning of 2016. The inclusion of VIX as a covariate seems to better capture the highest peaks of volatility, but it also seems to overestimate volatility at other points. The GARCH model and the ESVM seem to give very similar estimates, however the ESVM seems to be able to transition from high volatility to low volatility more quickly. This is likely due to the positive parameter restrictions required for GARCH.

In addition to visual comparison of these volatility models, we also use DIC as recommended by \citet{berg2004dic} and WAIC for model comparison of stochastic volatility models. The values of these information criteria under each approach can be found in Table~\ref{tab: vol}. We find that both ESVM approaches improve upon the GARCH model in terms of DIC. In terms of WAIC, the base ESVM is superior, followed by the GARCH model.  Surprisingly, the addition of VIX as a covariate does not result in a better model fit than the ESVM without covariates. This is likely because the addition of VIX can lead to overestimates of volatility for many time periods. In fact, \citet{kownatzki2016good} show that VIX consistently overestimates volatility, when compared to actual realized volatility, during periods of normal volatility. In this case, the ESVM seems to adequately capture volatility without the use of covariates, though it may be desirable to find more appropriate covariates depending on the application. 

\begin{table}[H]
\begin{center}
 \begin{tabular}{||c | c c ||} 
 \hline
 Model & DIC & WAIC   \\ [0.5ex] 
 \hline\hline
 GARCH(1,1) & $-5.42 \times 10^{-3}$  & $-5.41 \times 10^{-3}$ \\ 
 \hline
 ESVM & $\bm{-5.65 \times 10^{-3}}$ & $\bm{-5.60 \times 10^{-3}}$ \\
 \hline
 ESVM with Covariate & $-5.44 \times 10^{-3}$ & $-5.17 \times 10^{-3}$  \\ [1ex] 
 \hline
\end{tabular}
\caption{DIC and WAIC values for each of the three volatility models using Dow Jones data.}
\label{tab: vol}
\end{center}
\end{table}

\begin{figure}[H]
    \begin{center}
        \includegraphics[width=150mm]{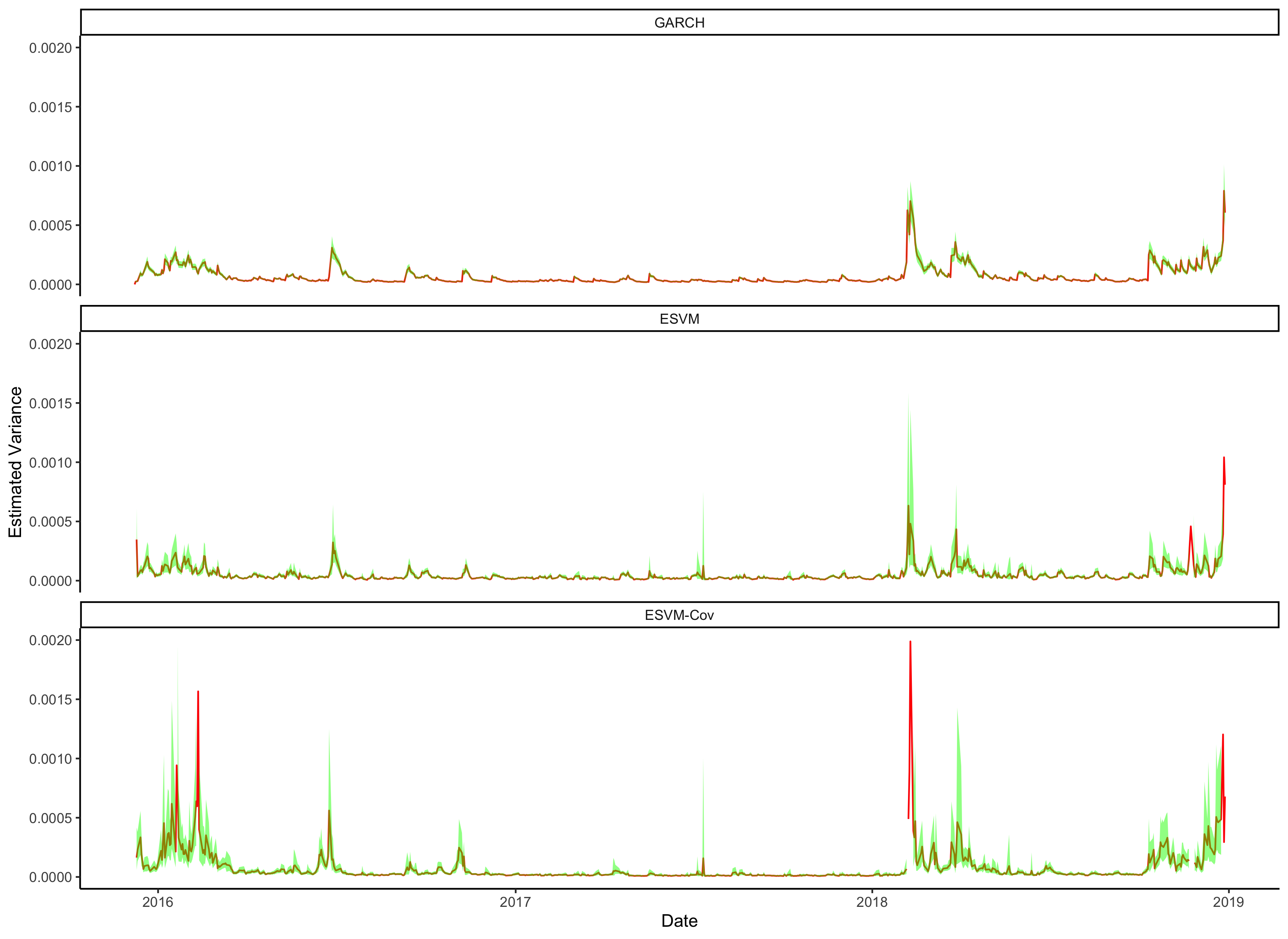}
         \caption{\baselineskip=10pt Volatility estimates (red) and accompanying pointwise 95\% credible intervals (green) under the three models fit to the Dow Jones data.}
         \label{fig: ci}
    \end{center}               
\end{figure}

Lastly, we emphasize that as a byproduct of fitting these models in a Bayesian setting, we attain uncertainty quantification. We illustrate this via Figure \ref{fig: ci}, which shows the volatility estimates as well as pointiwse 95\% credible intervals. In each case, uncertainty is greatest during periods of high volatility. The ESVM has slightly more uncertainty than the GARCH model. The addition of VIX as a covariate adds additional uncertainty. The fact that the ESVM yields uncertainty quantification is especially important, as it uses a neural network architecture. Typically neural networks are fit via stochastic gradient descent, which does not necessarily permit estimates of uncertainty.

\section{Creatinine Analysis}\label{sec: creatinine}

As a final example, we consider the creatinine dataset used by \citep{liu1995ml}. This data consists of clinical trial measurements of endogenous creatinine clearance (CR) for 34 male patients. We also consider age and serum creatinine (SC) concentration as covariates. We remove observations without a measurement of SC, leaving 30 observations for analysis. This dataset exhibits heavy tails, and thus motivates the use of the Laplace BGHM.

We fit four different models to the data:
\begin{enumerate}
    \item \textbf{Model 1} uses the Gaussian BGHM with constant variance by letting $\bm{x}_{1i}$ contain an intercept as well as age and SC, while $\bm{x}_{2i}$ only contains an intercept.
    \item \textbf{Model 2} uses the Gaussian BGHM with modeled variance by letting $\bm{x}_{1i}$ contain an intercept as well as age and SC, while $\bm{x}_{2i}$ contains and intercept and SC.
    \item \textbf{Model 3} uses the Laplace BGHM with constant variance by letting $\bm{x}_{1i}$ contain an intercept as well as age and SC, while $\bm{x}_{2i}$ only contains an intercept.
    \item \textbf{Model 4} uses the Laplace BGHM with modeled variance by letting $\bm{x}_{1i}$ contain an intercept as well as age and SC, while $\bm{x}_{2i}$ contains and intercept and SC.
\end{enumerate}
In every case, the vectors $\bm{\Psi}_{1i}$ and $\bm{\Psi}_{2i}$ are empty. We present DIC and WAIC from fitting the models to the data in Table \ref{tab: creatinine}. There is clear indication that the conditionally Laplace data models fit better than the conditionally Gaussian models. However, it also appears that the use of SC as a covariate with respect to the variance may be important regardless of which data model is used. 

\begin{table}[H]
\begin{center}
 \begin{tabular}{||c | c | c||} 
 \hline
 Model & DIC & WAIC  \\ [0.5ex] 
 \hline\hline
 Model 1 & $21.9$  & $78.6$ \\ 
 \hline
 Model 2 & $16.7$ & $114.5$ \\
 \hline
 Model 3 & $3.3$  & $8.5$ \\ 
 \hline
  Model 4 & $2.0$  & $8.5$ \\ [1ex] 
  \hline
\end{tabular}
\caption{DIC and WAIC values for each of the models fit to the creatinine data.}
\label{tab: creatinine}
\end{center}
\end{table}

\section{Discussion}\label{sec: disc}

Our main contribution in this work is twofold. First, we have constructed a general model for heteroskedastic data that can be used in a variety of settings. Covariates (discrete or continuous) can easily be used to flexibly model the conditional variance. Furthermore, the mixed model approach allows for basis function approaches to be used, which can model various dependencies in the data. Second, we have added the appropriate link function and prior distribution to allow for conjugate sampling of the full conditional distribution in a Bayesian setting. Thus, our model is not only general, but also straightforward to fit, without requiring any tuning. In addition to these contributions, we have also extended the model to a deep learning setting through the use of an echo state network architecture. This extension allows for highly flexible nonlinear modeling, especially in the context of time depenedent data. Finally, we provide further extension to heavy-tailed data through the use of a Laplace data model. We envision many more extensions being recognized as this methodology becomes used for various applications.

Although not explored in this work, it would be interesting to compare prediction intervals under different models for heteroskedastic data. Intuitively, we would expect better coverage rates for models that better specify the conditional variance function. Thus, we would expect the generality of the GBHM to allow for better coverage rates than models that assume constant variance or have limited modeling of the variance.

Finally, future work may include exploration of other nonlinear modeling approaches within this framework. For example, similar networks to the ESN exist for the case of data without temporal dependence. It would be straightforward to incorporate these into the GBHM. In addition, the asymptotic relationship between the MLG and Gaussian distribution may allow for some type of Gaussian process approach to be used within the GBHM. Another avenue of future work could involve exploration of other heavy tailed data models. We used a Laplace data model to account for heavy tail phenomena, but other distributions such as Cauchy may prove valuable as well. Further data augmentation strategies may be available in these cases to retain conjugate sampling. Lastly, with respect to the ESVM, it may be desirable to consider more hidden layers. \citet{mcdermott2019deep} found some benefit to modeling the mean function with multiple hidden layers, giving reason to believe that it may be valuable when modeling the variance as well.

\section*{Acknowledgements}
Support for this research through the Census Bureau Dissertation Fellowship program is gratefully acknowledged.  This research was partially supported by the U.S. National Science Foundation (NSF) under NSF SES-1853096 and through the Air Force Research Laboratory (AFRL) Contract No.19C0067.  The views expressed on statistical issues are those of the authors and not those of the NSF or the U.S. Air Force. This work was supported in part by the U.S. Department of Agriculture, Natural Resources Conservation Service. The findings and conclusions in this publication are those of the authors and should not be construed to represent any official USDA or U.S. Government determination or policy.

\bibliography{bghm}
\bibliographystyle{jasa}

\section*{Appendix A: Full conditional distributions for the BGHM}

Let $\bm{\Sigma}_y=\hbox{Diag}(1/\sigma^2_1,\ldots,1/\sigma^2_n)$ and $\bm{\mu}_y=(\mu_1,\ldots,\mu_n)$.

\begin{align*}
    \bm{\beta}_1 | \cdot & \propto \hbox{exp}\left(-\frac{1}{2}(\bm{y} - \bm{X}_1 \bm{\beta}_1 - \bm{\Psi}_1 \bm{\eta}_1)'\bm{\Sigma}_y(\bm{y} - \bm{X}_1 \bm{\beta}_1 - \bm{\Psi}_1 \bm{\eta}_1) \right) \\
    & \times \hbox{exp}\left(-\frac{1}{2\sigma^2_{\beta_1}} \bm{\beta}'\bm{\beta} \right) \\
    \bm{\beta}_1 | \cdot & \sim \hbox{N}_{p_1}\left( 
        \bm{\mu}=(\bm{X}_1'\bm{\Sigma}_y\bm{X}_1 + \frac{1}{\sigma^2_{\beta_1}} \bm{I}_{p_1})^{-1} \bm{X}_1'\bm{\Sigma}_y(\bm{y} - \bm{\Psi}_1\bm{\eta}_1),
        \bm{\Sigma} = (\bm{X}_1'\bm{\Sigma}_y\bm{X}_1 + \frac{1}{\sigma^2_{\beta_1}} \bm{I}_{p_1})^{-1}
    \right) 
\end{align*}
    
\begin{align*}
    \bm{\eta}_1 | \cdot & \propto \hbox{exp}\left(-\frac{1}{2}(\bm{y} - \bm{X}_1 \bm{\beta}_1 - \bm{\Psi}_1 \bm{\eta}_1)'\bm{\Sigma}_y(\bm{y} - \bm{X}_1 \bm{\beta}_1 - \bm{\Psi}_1 \bm{\eta}_1) \right) \\
    & \times \hbox{exp}\left(-\frac{1}{2\sigma^2_{\eta_1}} \bm{\eta}'\bm{\eta} \right) \\
    \bm{\eta}_1 | \cdot & \sim \hbox{N}_{r_1}\left( 
        \bm{\mu}=(\bm{\Psi}_1'\bm{\Sigma}_y\bm{\Psi}_1 + \frac{1}{\sigma^2_{\eta_1}} \bm{I}_{r_1})^{-1} \bm{\Psi}_1'\bm{\Sigma}_y(\bm{y} - \bm{X}_1\bm{\beta}_1),
        \bm{\Sigma} = (\bm{\Psi}_1'\bm{\Sigma}_y\bm{\Psi}_1 + \frac{1}{\sigma^2_{\eta_1}} \bm{I}_{r_1})^{-1}
    \right) 
\end{align*}

\begin{align*}
    \bm{\beta}_2 | \cdot & \propto \prod_{i=1}^n \hbox{exp} \left\{\frac{1}{2}\bm{x}_{2i}'\bm{\beta}_2 - \frac{1}{2}(y_i - \mu_i)^2\hbox{exp}(\bm{\psi}_{2i}'\bm{\eta}_2)\hbox{exp}(\bm{x}_{2i}'\bm{\beta}_2) \right\} \\
    & \times \hbox{exp}\left\{\alpha \bm{1}_{p_2}' \alpha^{-1/2} \frac{1}{\sigma_{\beta_2}} \bm{I}_{p_2} \bm{\beta}_2 - \alpha \bm{1}_{p_2}' \hbox{exp}\left(\alpha^{-1/2} \frac{1}{\sigma_{\beta_2}} \bm{I}_{p_2} \bm{\beta}_2\right)  \right\} \\
            &= \hbox{exp}\left\{\bm{\alpha}_{\beta_2}' \bm{H}_{\beta_2} \bm{\beta}_2 - \bm{\kappa}_{\beta_2}' \hbox{exp}(\bm{H}_{\beta_2} \bm{\beta}_2)\right\} \\
        & \bm{H}_{\beta_2} = \left[
                            \begin{array}{c}
                            \bm{X}_2  \\
                            \alpha^{-1/2} \frac{1}{\sigma_{\beta_2}} \bm{I}_{p_2} 
                            \end{array}
                            \right], \quad 
        \bm{\alpha}_{\beta_2} = \left(\frac{1}{2} \bm{1}_n', \alpha \bm{1}_{p_2}'\right)', \quad \\
        & \bm{\kappa}_{\beta_2} = \left(\frac{1}{2}\left\{(\bm{y}-\bm{\mu}_y)^2 \odot \hbox{exp}(\bm{\Psi}_2 \bm{\eta}_2)\right\}', \alpha \bm{1}_{p_2}'\right)' \\
        \bm{\beta}_2 | \cdot &\sim \hbox{cMLG}(\bm{H}_{\beta_2}, \bm{\alpha}_{\beta_2}, \bm{\kappa}_{\beta_2})
\end{align*}

\begin{align*}
    \bm{\eta}_2 | \cdot & \propto \prod_{i=1}^n \hbox{exp} \left\{\frac{1}{2}\bm{\psi}_{2i}'\bm{\eta}_2 - \frac{1}{2}(y_i - \mu_i)^2\hbox{exp}(\bm{x}_{2i}'\bm{\beta}_2)\hbox{exp}(\bm{\psi}_{2i}'\bm{\eta}_2) \right\} \\
    & \times \hbox{exp}\left\{\alpha \bm{1}_{r_2}' \alpha^{-1/2} \frac{1}{\sigma_{\eta_2}} \bm{I}_{r_2} \bm{\eta}_2 - \alpha \bm{1}_{r_2}' \hbox{exp}\left(\alpha^{-1/2} \frac{1}{\sigma_{\eta_2}} \bm{I}_{r_2} \bm{\eta}_2\right)  \right\} \\
            &= \hbox{exp}\left\{\bm{\alpha}_{\eta_2}' \bm{H}_{\eta_2} \bm{\eta}_2 - \bm{\kappa}_{\eta_2}' \hbox{exp}(\bm{H}_{\eta_2} \bm{\eta}_2)\right\} \\
        & \bm{H}_{\eta_2} = \left[
                            \begin{array}{c}
                            \bm{\Psi}_2  \\
                            \alpha^{-1/2} \frac{1}{\sigma_{\eta_2}} \bm{I}_{r_2} 
                            \end{array}
                            \right], \quad 
        \bm{\alpha}_{\eta_2} = \left(\frac{1}{2} \bm{1}_n', \alpha \bm{1}_{r_2}'\right)', \quad \\
        & \bm{\kappa}_{\eta_2} = \left(\frac{1}{2}\left\{(\bm{y}-\bm{\mu}_y)^2 \odot \hbox{exp}(\bm{X}_2 \bm{\beta}_2)\right\}', \alpha \bm{1}_{r_2}'\right)' \\
        \bm{\eta}_2 | \cdot &\sim \hbox{cMLG}(\bm{H}_{\eta_2}, \bm{\alpha}_{\eta_2}, \bm{\kappa}_{\eta_2})
\end{align*}

\begin{align*}
    \sigma^2_{\eta_1} | \cdot & \propto \left(\sigma^2_{\eta_1}\right)^{-r_1/2} \hbox{exp}\left(-\frac{1}{2 \sigma^2_{\eta_1}} \bm{\eta}_1'\bm{\eta}_1 \right) \times \left(\sigma^2_{\eta_2}\right)^{-a -1}\hbox{exp}\left(-\frac{b}{\sigma^2_{\eta_1}}\right) \\ 
    &= \left(\sigma^2_{\eta_1}\right)^{-(a + r_1/2) - 1} \hbox{exp}\left\{-\frac{1}{\sigma^2_{\eta_2}}\left(b + \frac{\bm{\eta}_1'\bm{\eta}_1}{2}\right) \right\} \\
    \sigma^2_{\eta_1} | \cdot & \sim \hbox{IG}\left(a + \frac{r_1}{2}, \; b + \frac{\bm{\eta}_1'\bm{\eta}_1}{2} \right)
\end{align*}

\begin{align*}
    \frac{1}{\sigma_{\eta_2}} | \cdot & \propto \hbox{exp}\left\{\alpha \bm{1}_{r_2}' \alpha^{-1/2} \frac{1}{\sigma_{\eta_2}} \bm{I}_{r_2} \bm{\eta}_2 - \alpha \bm{1}_{r_2}' \hbox{exp}\left(\alpha^{-1/2} \frac{1}{\sigma_{\eta_2}} \bm{I}_{r_2} \bm{\eta}_2\right)  \right\} \\
        & \quad \times \hbox{exp}\left\{\omega \frac{1}{\sigma_{\eta_2}} - \rho \, \hbox{exp}\left(\frac{1}{\sigma_{\eta_2}}\right)\right\} \times I(\sigma_{\eta_2} > 0) \\
                & = \hbox{exp} \left\{\bm{\omega}_{\sigma}' \bm{H}_{\sigma} \frac{1}{\sigma_{\eta_2}} - \bm{\rho}_{\sigma}' \hbox{exp} \left( \bm{H}_{\sigma} \frac{1}{\sigma_{\eta_2}} \right) \right\} \times I(\sigma_{\eta_2} > 0) \\
        & \bm{H}_{\sigma} = (\alpha^{-1/2} \bm{\eta}_2', 1)' \quad
        \bm{\omega}_{\sigma} = (\alpha \bm{1}_{r_2}', \omega)' \quad 
        \bm{\rho}_{\sigma} = (\alpha \bm{1}_{r_2}', \rho)'\\
        \frac{1}{\sigma_{\eta_2}} | \cdot & \sim \hbox{cMLG}(\bm{H}_{\sigma}, \bm{\omega}_{\sigma}, \bm{\rho}_{\sigma}) \times I(\sigma_{\eta_2} > 0) 
\end{align*}

\section*{Appendix B: Full conditional distributions for the Laplace BGHM}
Let $\bm{\Sigma}_y=\hbox{Diag}(1/s_1,\ldots,1/s_n)$ and $\bm{\mu}_y=(\mu_1,\ldots,\mu_n)$.

\begin{align*}
    \frac{1}{s_i} | \cdot & \propto \left(\frac{1}{s_i}\right)^{-3/2} \hbox{exp}\left\{-\frac{1}{2}\left(\frac{1}{s_i}\right)(y_i - \mu_i)^2  \right\} \\
    & \times \hbox{exp}\left\{-\frac{1}{\sigma^2_i} \left(\frac{1}{s_i} \right)^{-1} \right\} I(s_i > 0) \\
    & \propto \left(\frac{1}{s_i}\right)^{-3/2} \hbox{exp}\left\{- \frac{(y_i - \mu_i)^2 \left(\frac{1}{s_i} - \left\{\frac{2}{(y_i-\mu_i)^2\sigma^2_i} \right\}^{1/2}\right)^2}{2 \frac{1}{s_i}}  \right\}I(s_i > 0) \\
    \frac{1}{s_i} | \cdot & \sim \hbox{Inverse Gaussian}\left(\mu_s = \left\{\frac{2}{(y_i-\mu_i)^2\sigma^2_i} \right\}^{1/2}, \; \lambda_s=\frac{2}{\sigma^2_i} \right) 
\end{align*}

\begin{align*}
    \bm{\beta}_1 | \cdot & \propto \hbox{exp}\left(-\frac{1}{2}(\bm{y} - \bm{X}_1 \bm{\beta}_1 - \bm{\Psi}_1 \bm{\eta}_1)'\bm{\Sigma}_y(\bm{y} - \bm{X}_1 \bm{\beta}_1 - \bm{\Psi}_1 \bm{\eta}_1) \right) \\
    & \times \hbox{exp}\left(-\frac{1}{2\sigma^2_{\beta_1}} \bm{\beta}'\bm{\beta} \right) \\
    \bm{\beta}_1 | \cdot & \sim \hbox{N}_{p_1}\left( 
        \bm{\mu}=(\bm{X}_1'\bm{\Sigma}_y\bm{X}_1 + \frac{1}{\sigma^2_{\beta_1}} \bm{I}_{p_1})^{-1} \bm{X}_1'\bm{\Sigma}_y(\bm{y} - \bm{\Psi}_1\bm{\eta}_1),
        \bm{\Sigma} = (\bm{X}_1'\bm{\Sigma}_y\bm{X}_1 + \frac{1}{\sigma^2_{\beta_1}} \bm{I}_{p_1})^{-1}
    \right) 
\end{align*}
    
\begin{align*}
    \bm{\eta}_1 | \cdot & \propto \hbox{exp}\left(-\frac{1}{2}(\bm{y} - \bm{X}_1 \bm{\beta}_1 - \bm{\Psi}_1 \bm{\eta}_1)'\bm{\Sigma}_y(\bm{y} - \bm{X}_1 \bm{\beta}_1 - \bm{\Psi}_1 \bm{\eta}_1) \right) \\
    & \times \hbox{exp}\left(-\frac{1}{2\sigma^2_{\eta_1}} \bm{\eta}'\bm{\eta} \right) \\
    \bm{\eta}_1 | \cdot & \sim \hbox{N}_{r_1}\left( 
        \bm{\mu}=(\bm{\Psi}_1'\bm{\Sigma}_y\bm{\Psi}_1 + \frac{1}{\sigma^2_{\eta_1}} \bm{I}_{r_1})^{-1} \bm{\Psi}_1'\bm{\Sigma}_y(\bm{y} - \bm{X}_1\bm{\beta}_1),
        \bm{\Sigma} = (\bm{\Psi}_1'\bm{\Sigma}_y\bm{\Psi}_1 + \frac{1}{\sigma^2_{\eta_1}} \bm{I}_{r_1})^{-1}
    \right) 
\end{align*}

\begin{align*}
    \bm{\beta}_2 | \cdot & \propto \prod_{i=1}^n \hbox{exp} \left\{\bm{x}_{2i}'\bm{\beta}_2 - s_i\hbox{exp}(\bm{\psi}_{2i}'\bm{\eta}_2)\hbox{exp}(\bm{x}_{2i}'\bm{\beta}_2) \right\} \\
    & \times \hbox{exp}\left\{\alpha \bm{1}_{p_2}' \alpha^{-1/2} \frac{1}{\sigma_{\beta_2}} \bm{I}_{p_2} \bm{\beta}_2 - \alpha \bm{1}_{p_2}' \hbox{exp}\left(\alpha^{-1/2} \frac{1}{\sigma_{\beta_2}} \bm{I}_{p_2} \bm{\beta}_2\right)  \right\} \\
            &= \hbox{exp}\left\{\bm{\alpha}_{\beta_2}' \bm{H}_{\beta_2} \bm{\beta}_2 - \bm{\kappa}_{\beta_2}' \hbox{exp}(\bm{H}_{\beta_2} \bm{\beta}_2)\right\} \\
        & \bm{H}_{\beta_2} = \left[
                            \begin{array}{c}
                            \bm{X}_2  \\
                            \alpha^{-1/2} \frac{1}{\sigma_{\beta_2}} \bm{I}_{p_2} 
                            \end{array}
                            \right], \quad 
        \bm{\alpha}_{\beta_2} = \left( \bm{1}_n', \alpha \bm{1}_{p_2}'\right)', \quad \\
        & \bm{\kappa}_{\beta_2} = \left(\left\{\bm{s} \odot \hbox{exp}(\bm{\Psi}_2 \bm{\eta}_2)\right\}', \alpha \bm{1}_{p_2}'\right)' \\
        \bm{\beta}_2 | \cdot &\sim \hbox{cMLG}(\bm{H}_{\beta_2}, \bm{\alpha}_{\beta_2}, \bm{\kappa}_{\beta_2})
\end{align*}

\begin{align*}
    \bm{\eta}_2 | \cdot & \propto \prod_{i=1}^n \hbox{exp} \left\{\bm{\psi}_{2i}'\bm{\eta}_2 - s_i\hbox{exp}(\bm{x}_{2i}'\bm{\beta}_2)\hbox{exp}(\bm{\psi}_{2i}'\bm{\eta}_2) \right\} \\
    & \times \hbox{exp}\left\{\alpha \bm{1}_{r_2}' \alpha^{-1/2} \frac{1}{\sigma_{\eta_2}} \bm{I}_{r_2} \bm{\eta}_2 - \alpha \bm{1}_{r_2}' \hbox{exp}\left(\alpha^{-1/2} \frac{1}{\sigma_{\eta_2}} \bm{I}_{r_2} \bm{\eta}_2\right)  \right\} \\
            &= \hbox{exp}\left\{\bm{\alpha}_{\eta_2}' \bm{H}_{\eta_2} \bm{\eta}_2 - \bm{\kappa}_{\eta_2}' \hbox{exp}(\bm{H}_{\eta_2} \bm{\eta}_2)\right\} \\
        & \bm{H}_{\eta_2} = \left[
                            \begin{array}{c}
                            \bm{\Psi}_2  \\
                            \alpha^{-1/2} \frac{1}{\sigma_{\eta_2}} \bm{I}_{r_2} 
                            \end{array}
                            \right], \quad 
        \bm{\alpha}_{\eta_2} = \left( \bm{1}_n', \alpha \bm{1}_{r_2}'\right)', \quad \\
        & \bm{\kappa}_{\eta_2} = \left(\left\{\bm{s} \odot \hbox{exp}(\bm{X}_2 \bm{\beta}_2)\right\}', \alpha \bm{1}_{r_2}'\right)' \\
        \bm{\eta}_2 | \cdot &\sim \hbox{cMLG}(\bm{H}_{\eta_2}, \bm{\alpha}_{\eta_2}, \bm{\kappa}_{\eta_2})
\end{align*}

\begin{align*}
    \sigma^2_{\eta_1} | \cdot & \propto \left(\sigma^2_{\eta_1}\right)^{-r_1/2} \hbox{exp}\left(-\frac{1}{2 \sigma^2_{\eta_1}} \bm{\eta}_1'\bm{\eta}_1 \right) \times \left(\sigma^2_{\eta_2}\right)^{-a -1}\hbox{exp}\left(-\frac{b}{\sigma^2_{\eta_1}}\right) \\ 
    &= \left(\sigma^2_{\eta_1}\right)^{-(a + r_1/2) - 1} \hbox{exp}\left\{-\frac{1}{\sigma^2_{\eta_2}}\left(b + \frac{\bm{\eta}_1'\bm{\eta}_1}{2}\right) \right\} \\
    \sigma^2_{\eta_1} | \cdot & \sim \hbox{IG}\left(a + \frac{r_1}{2}, \; b + \frac{\bm{\eta}_1'\bm{\eta}_1}{2} \right)
\end{align*}

\begin{align*}
    \frac{1}{\sigma_{\eta_2}} | \cdot & \propto \hbox{exp}\left\{\alpha \bm{1}_{r_2}' \alpha^{-1/2} \frac{1}{\sigma_{\eta_2}} \bm{I}_{r_2} \bm{\eta}_2 - \alpha \bm{1}_{r_2}' \hbox{exp}\left(\alpha^{-1/2} \frac{1}{\sigma_{\eta_2}} \bm{I}_{r_2} \bm{\eta}_2\right)  \right\} \\
        & \quad \times \hbox{exp}\left\{\omega \frac{1}{\sigma_{\eta_2}} - \rho \, \hbox{exp}\left(\frac{1}{\sigma_{\eta_2}}\right)\right\} \times I(\sigma_{\eta_2} > 0) \\
                & = \hbox{exp} \left\{\bm{\omega}_{\sigma}' \bm{H}_{\sigma} \frac{1}{\sigma_{\eta_2}} - \bm{\rho}_{\sigma}' \hbox{exp} \left( \bm{H}_{\sigma} \frac{1}{\sigma_{\eta_2}} \right) \right\} \times I(\sigma_{\eta_2} > 0) \\
        & \bm{H}_{\sigma} = (\alpha^{-1/2} \bm{\eta}_2', 1)' \quad
        \bm{\omega}_{\sigma} = (\alpha \bm{1}_{r_2}', \omega)' \quad 
        \bm{\rho}_{\sigma} = (\alpha \bm{1}_{r_2}', \rho)'\\
        \frac{1}{\sigma_{\eta_2}} | \cdot & \sim \hbox{cMLG}(\bm{H}_{\sigma}, \bm{\omega}_{\sigma}, \bm{\rho}_{\sigma}) \times I(\sigma_{\eta_2} > 0) 
\end{align*}

\end{document}